\documentclass[a4paper,10pt,twoside]{article}

\usepackage{graphics}

\usepackage{amsmath,amssymb,amsfonts}
\usepackage[dvips]{graphicx}
\usepackage{authblk}
\usepackage{multicol}
\usepackage{epstopdf}
\usepackage[hmargin=2.1cm,vmargin=1.5cm]{geometry}
\usepackage{float}
\usepackage[font=it]{caption}
\usepackage[english]{babel}
\usepackage[scaled]{helvet}
\usepackage[T1]{fontenc}
\usepackage{url}
\usepackage[colorlinks=true, allcolors=blue]{hyperref}
\usepackage[numbers,sort&compress]{natbib}

\title{A phase shift formulation for N-light-pulse atom interferometers: application to inertial sensing}

\author[1,2*]{Malo Cadoret}
\author[2]{Nassim Zahzam}
\author[2]{Yannick Bidel}
\author[2]{Cl\'ement Diboune}
\author[2]{Alexis Bonnin}
\author[2]{Fabien Th\'eron}
\author[2]{Alexandre Bresson}

\affil[1]{Laboratoire Commun de M\'{e}trologie, CNAM ,61 rue du Landy,93210 La Plaine Saint-Denis, France}
\affil[2]{ONERA--The French Aerospace Lab, F-91123 Palaiseau cedex, France}

\affil[*]{Corresponding author: malo.cadoret@lecnam.net}

\date{}
\begin{document}
\maketitle

\begin{abstract}
We report on an original and simple formulation of the phase shift in $N$-light-pulse atom interferometers. We consider atomic interferometers based on two-photon transitions (Raman transitions or Bragg pulses). Starting from the exact analytical phase shift formula obtained from the atom optics ABCD formalism, we use a power series expansion in time  of the position of the atomic wave packet with respect to the initial condition. The result of this expansion leads to a formulation of the interferometer phase shift where the leading coefficient in the phase terms up to $T^{k}$ dependences ($k\geq 0$) in the time separation $T$ between pulses, can be simply expressed in terms of a  product between a Vandermonde matrix, and a vector characterizing the two-photon pulse sequence of the interferometer.  
This simple coefficient dependence of the phase shift reflects very well the atom interferometer's sensitivity to a specific inertial field in the presence of multiple gravito-inertial effects. Consequently, we show that this formulation is well suited when looking for selective atomic sensors of accelerations, rotations or photon recoil only, which can be obtained by simply zeroing some specific coefficients. We give a theoritical application of our formulation to the photon recoil measurement.
\end{abstract}

\begin{multicols}{2}

\section{Introduction}
Since the first demonstration of cold atom interferometric inertial sensors \cite{Kasevich1991,Riehle1991} in the 1990s, light-pulse Atom Interferometers (AIs) have become very stable and extremely accurate sensors for the measurements of fundamental constants such as gravitational constant \cite{Rosi2014} or the fine structure constant \cite{Cadoret2008,Bouchendira2011}, and inertial forces like gravity acceleration \cite{Peters1999,Gouet2008}, Earth's gravity gradient \cite{McGuirk2002,Tino2014} or rotations \cite{Gustavson1997,Stockton2011,Landragin2016} finding variety of applications in geodesy, geophysics, metrology, inertial navigation among others. In addition, high sensitivity AIs have become promising candidates for laboratory tests of General Relativity (GR) \cite{Dimopoulos2007} especially for gravitational-waves (GW) detection \cite{Dimopoulos2008,Chaibi2016}, short-range forces \cite{Wolf2007} and tests of the weak equivalence principle (WEP) \cite{Bonnin2013}. For all applications, high resolution measurements must take into account  high order terms in the phase shift to increase the sensors accuracy and sensitivity. \newline

Phase shift calculations in light-pulse AIs in the presence of multiple gravito-inertial fields have already been deeply investigated \cite{Bongs2006,Dubetsky2006,Borde1992,Borde2002}. However, the phase shift formulations are not always convenient when one wants to use them for practical applications  such as the development of atomic sensors that could be selectively sensitive to a particular inertial field such as a rotation, acceleration as well as cross terms. In this letter we derive a novel formulation for phase shift calculations in $N$ light-pulse AIs considering the case of two-path interferences in the presence of multiple gravito-inertial fields. The formulation presented in this letter  could be of practical interest both for the understanding of contribution of inertial terms to the phase shift, as well as for applications to the development of selective inertial sensors. This simple and exact formulation is obtained starting from an exact analytical phase shift formula valid for any pulse sequence and taking into account gravity, gravity gradient and rotations \cite{Antoine2003}.\newline
The article is organized as follows.
In section~\ref{sec:calculations} we introduce the general framework of our calculations and give a simple formulation of the phase shift up to $T^k$ dependences. We explicitely derive the phase shift up to order $k=4$ in the presence of multiple gravito-inertial fields. In section~\ref{sec:analysis} we give a physical analysis of the leading coefficients of the phase shift obtained with our formulation. In section~\ref{sec:photonrecoil} we consider the effects of wave vector change and photon recoil on the atomic phase shift in the presence of inertial forces. In section~\ref{sec:application} we  give the expression of the total phase shift for any N-light pulse AI according to our formulation. We then apply our formula to the case of a Mach-Zehnder atomic interferometer and evaluate the phase shift terms of our compact cold rubidium atom gravimeter.  We show that our results are consistent with previous work of Antoine et al., Dubetsky et al., and Wolf et al. \cite{Antoine2003,Dubetsky2006,Wolf1999}. In section~\ref{sec:originalai} we demonstrate the benefit of our formulation  when searching for a particular selective inertial atomic sensor. We give theoritical examples of N-light-pulse AIs dedicated to the photon recoil measurement.
\section{Calculations}
\label{sec:calculations}
\subsection{Notations}

We consider $N$-light-pulse AIs in which atoms undergo two-photon transitions (Raman transition or Bragg pulse) and where the two laser beams are counter-propagating or co-propagating with pulse sequences only separated in time. We consider AIs in the limit of short pulses.
A two-photon transition is treated as a single photon transition with effective frequency $\omega=\omega_{1}-\omega_{2}$ and effective wave vector $k=|\vec{k}_{1}-\vec{k}_{2}|$ corresponding to the difference in frequency and wave vector respectively. In our formalism, $2M$-photon Bragg transitions (with $M$ being the Bragg diffraction order, $M \geq 1$), could be considered by simply giving to the effective wave vector a magnitude $Mk$. In the following calculations, we will consider $M=1$, unless specified. Thus, a Bragg transition will be equivalent to a Raman transition if one does not matter about the quantum internal state. We will precise wether the internal quantum state changes (Raman pulse) or not (Bragg pulse) if necessary.
\subsubsection{Time and laser field notations}
We first consider AIs for which light pulses are all equally separated by time $T$,(extension to the case of an arbitrary pulse sequence is possible (see section 3~\ref{sec:arbitrarypulse})).
The AIs are cut into as many slices as there are interactions.\newline
Hence, the time of the $i$-th light-pulse is defined as :
\begin{equation}
t_{i}=(i-1)T\,\,\,\,\,\,\,\,\,\,\,\,\,\,(i\geq 1)
\label{eq:temps}
\end{equation}
The effective frequencies and wave vectors will be considered different for each pulse in order to compensate for Doppler shift. Hence, the effective laser frequency is chirped linearly so as to maintain the resonance condition for each light-pulse as it is in the case of gravity, and resonance condition, will be satisfied.\newline
The effective wave vector $k_{i}$ and frequency $\omega_{i}$ of the two-photon transition of the $i$-th light pulse are defined as:$\\$
\begin{center}
	\begin{equation}
		k_{i}=k_1+\Delta k(i-1)
	\end{equation}
\end{center}
\begin{equation}
	\omega_{i}=\omega_1 +\Delta\omega (i-1)
\end{equation}
Where $\Delta k$  accounts for the difference in effective wave-vector and $\Delta \omega$ is  the frequency difference due to the laser chirp, independent of the laser pulse number $i$. For instance, $\Delta k$ is equal to zero when chirping symmetrically two counter propagative laser beams, whereas $\Delta \omega$ is not.

We introduce an effective laser phase:
\begin{equation}
\phi_{i}=\phi_{1}+\omega_1 (i-1)T+\Delta \omega(i-1)^{2}\frac{T}{2}
\label{eq:defphase}
\end{equation}

\subsubsection{Interferometric variables}
\begin{figure}[H]
\centering
\includegraphics[width=\linewidth]{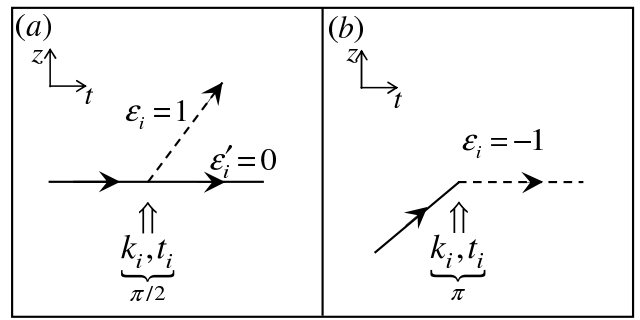}
\caption{Recoil diagram of a two-photon transition: the two-photon transition (Raman or Bragg) couples two momentum states. During the process the momentum transfer $\Delta p_i=\epsilon_i\hbar k_i$ at time $t_i$; (a) Beam-splitter pulse : the atom is put in a coherent superposition of two momentum states; (b) Mirror pulse: the atom's momentum state is changed with momentum $\Delta p$. Solid and dashed lines correspond to two different momentum states correponding to the same internal quantum state (Bragg pulse) or different internal quantum states (Raman transition). The two paths are described with vector $(\vec{\epsilon})$ and $(\vec{\epsilon'})$.}
\label{fig1}
\end{figure}

Any of the AI's that we consider will consist of time sequences of two kinds of pulses \cite{Berman1996}:\newline
\begin{itemize}
\item $\pi/2$ pulses that will play the role of matter-wave beam splitters.
 \item$\pi$ pulses that will act as matter-wave reflectors (or mirrors).
\end{itemize}
One path of an $N$-pulse AI is described by a vector $\vec{\epsilon}=(\epsilon_1,\epsilon_2,\cdots,\epsilon_i,\cdots,\epsilon_N)$ where $\epsilon_i$ accounts for the angular splitting induced by the atom-light interaction at time $t_{i}$, with  $\epsilon_i=+1$ and $\epsilon_i=-1$ for an upward and downward  momentum transfer $\Delta p_i=\epsilon_i\hbar k_i$ respectively, and where $k_i$ is along the $z$ direction. When the atom remains in the same momentum state, $\epsilon_i=0$. An example of pulse beam-splitter (i.e. $\pi/2$ pulse) and mirror pulse (i.e. $\pi$ pulse) are given in Fig.~\ref{fig1}.

\begin{figure}[H]
\begin{center}
\includegraphics[width=\linewidth,height=6cm]{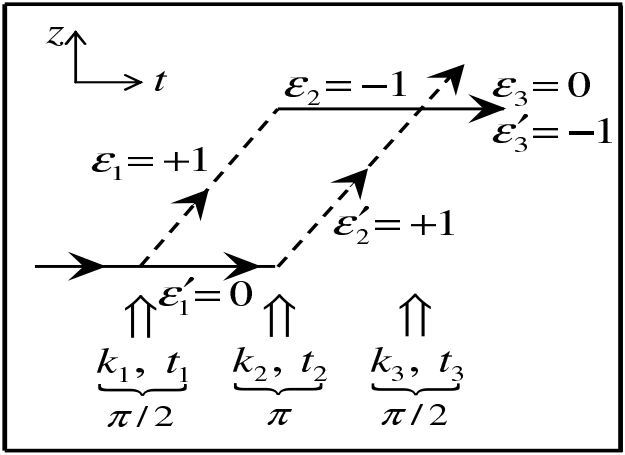}
\caption{Space-time recoil diagram in the absence of gravity of a Mach-Zehnder interferometer consisting of pulse-sequence $\pi/2-\pi-\pi/2$.}
\label{fig2}
\end{center}
\end{figure}

We will consider two-path AIs. Each path is described by a vector $G_k$ (upper path) and $G'_k$ (lower path) which takes into account the two-photon pulse sequence:
\begin{eqnarray}
\begin{split}
G_k=&\sum_{i=1}^{N} (i-1)^{k}\epsilon_{i}\\
G'_k=&\sum_{i=1}^{N} (i-1)^{k}\epsilon'_{i}
\end{split}
\end{eqnarray}
with $k$ an integer ($k\geq 0$).\newline
We introduce:
\begin{eqnarray}
\begin{split}
\Delta G_{k}=G_k-G'_k &=\sum_{i=1}^{N} (i-1)^{k}\Delta \epsilon_{i}\\
&=V\Delta \epsilon_{i}
\label{eq:deltaG}
\end{split}
\end{eqnarray}
 where $V$ is a Vandermonde matrix with coefficient $v_{mn}=(n-1)^{m-1}$ and where $\Delta \epsilon_i=\epsilon_{i}-\epsilon'_{i}$ is the difference of interaction between two paths at time $t_{i}$. 
As an example, we give in Fig.~\ref{fig2} the case of the well-known two-path Mach-Zehnder atomic interferometer, and focusing on one exit port of the interferometer one gets for the upper path $\vec{\epsilon}=(1,-1,0)$ and $\vec{\epsilon'}=(0,1,-1)$. Consequently one finds $\Delta \vec{\epsilon}=(1,-2,1)$. One could have chosen the other exit port of the interferometer which would not affect $\Delta \vec{\epsilon}$.

\subsection{Atom Interferometer phase shift calculation}
The phase shift in a light-pulse AI is often calculated in three steps. In a first step one identifies the different arms of the interferometer and calculates the phase shift due to atom-light interaction and free propagation of the atoms along each independent arm. In a second step one makes the difference between the phase shift obtained in each arm. Finally, in a third step, a phase shift due to spatial separation of the atoms is added.
To calculate the total phase shift one can use the Feynman path integral formalism \cite{Storey}, considering atoms as plane waves.
One limitation of this approach is that it is difficult to perform when one wants to account for simultaneous gravito-inertial effects such as gravity, rotation, gravity gradient and their interplay, as addressed in this paper.\newline 
In this context, the starting point of our approach is based on the atom optics $ABCD$ formalism developed by Bord\'e \cite{Borde1992}. This method allows both to consider atoms as wave packets, as well as to calculate the exact interferometric phase shift arising from multiple inertial effects.
\subsubsection{N-light-pulse AI phase shift derivation}
For any time-dependent hamiltonian at most quadratic in position $z$ and momentum $p$, which is the case for essentially all atom interferometric applications, the exact phase shift formula for a $N$-light-pulse AI with wave vector $k$ along the $z=\frac{\vec{k}}{k}.\vec{R}$ direction is \cite{Antoine2003}:
\begin{equation}
\Delta\Phi_{\mathrm{Total}}=\underbrace{\sum_{i=1}^{N}k_{i}\Delta\epsilon_{i}\frac{z_{i}+z'_{i}}{2}}_{\Delta\Phi_{inertial}}+\underbrace{\sum_{i=1}^{N} \phi_{i}\Delta\epsilon_{i}}_{\Delta\Phi_t}
\label{eq:phaseBorde}
\end{equation}
The right hand side of equation (\ref{eq:phaseBorde}) has two separate contributions. The first summation term accounts for the inertial phase shift, that we will denote $\Delta\Phi_{\mathrm{inertial}}$ with $z_{i}$ (and $z'_{i}$) being the position of the wave packet at time $t_i$ of the upper path (and lower path) respectively, and where one assumes no offset in the initial positions of the wave packet, $\Delta z_1=z_1-z'_1=0$. This inertial phase shift depends on the classical mid-point position of the atoms at time $t_i$. The second summation term is a time-dependent laser phase shift that we will denote $\Delta\Phi_{t}$. 
According to ($\ref{eq:defphase}$) the time dependent laser phase shift for any AI consisting of $N$-light-pulse can be expressed as follows:
\begin{eqnarray}
\begin{split}
\Delta\Phi_{t}&=\sum_{i=1}^{N}\Delta\epsilon_{i}\left(\phi_{1}+\omega_1(i-1)T
+\Delta \omega(i-1)^{2}\frac{T}{2}\right)\\
&=\Delta G_{0}\phi_{1}+\Delta G_{1}\omega_1 T+\Delta G_{2}\Delta\omega\frac{T}{2}
\label{eq:phasetemporelle}
\end{split}
\end{eqnarray}
where $\Delta G_{0},\Delta G_{1},\Delta G_{2}$ are defined in ($\ref{eq:deltaG}$).\newline
Thus we will mainly focus on the inertial phase shift calculation.
For simplicity we will first assume no wave vector change ($\Delta k=0$), as well as no photon recoil effect. These high order terms will be taken into account in section~\ref{sec:photonrecoil}.
Thus, the absence of recoil effect leads to $z_i=z'_i$ in equation (\ref{eq:phaseBorde}) which can be written:
\begin{equation}
\Delta\Phi_{\mathrm{inertial}} \simeq\sum_{i=1}^{N}k_{i}\Delta\epsilon_{i}z_i
\label{eq:dephasageaprox}
\end{equation}
In our method, we calculate the inertial phase shift, by making a Taylor expansion in power of $T$ of the $z$ position of the wave packet with respect to the first light-pulse ($t_1=0$) of the interferometer, assuming $z_1=0$:
\begin{equation}
z_{i}=z(t_i)=\sum_{k=0}a_{k}t_{i}^{k}=\sum_{k=0}a_{k}T^{k}(i-1)^{k}
\end{equation}
where $a_{k}=\frac{1}{k!}\left(\frac{d^{k}z}{dt^{k}}\right)_{t=t_1}$.\newline
After some algebra the inertial phase shift is given by:
\begin{eqnarray}
\begin{split}
\Delta \Phi_{\mathrm{inertial}}&\simeq
\sum_{k=0}\Delta G_{k}k_1a_k T^{k}\\
&\simeq \sum_{k=0}\frac{\Delta G_{k}}{k!}k_1\left(\frac{d^{k}z}{dt^{k}}\right)_{t=t_1} T^{k}\\
\label{eq:dephasage}
\end{split}
\end{eqnarray}
where $\Delta G_{k}$ is defined in equation (\ref{eq:deltaG}).\newline
One can rewrite $(\ref{eq:dephasage})$ considering the action of any gravito-inertial fields on the atoms located at  position $\vec{R}$, leading to the final expression:
\begin{eqnarray}
\begin{split}
\Delta \Phi_{\mathrm{inertial}}\simeq&\sum_{k}\frac{\Delta G_k}{k!} T^k\\
&\times \vec{k}_1\cdot\vec{\mathcal{F}}_k\left(\vec{R}_1,\vec{v}_1,+ inertial\,forces\right)\\
\end{split}
\label{eq:dephasage2}
\end{eqnarray}
where we introduce as a notation the function $\vec{\mathcal{F}}_k=\left(\frac{d^{k}\vec{R}}{dt^{k}}\right)_{t=t_1}$. This function depends on the atomic initial position $\vec{R}_{1}$ and velocity $\vec{v}_1$, and any gravito-inertial forces experienced by the atoms. Moreover, it is independent of the AI geometry.  We will apply this novel formulation of the phase shift obtained in equation (\ref{eq:dephasage2}) in the next section.\newline
\subsection{Phase shift due to gravito-inertial fields}
\begin{figure}[H]
\begin{center}
\includegraphics[width=\linewidth]{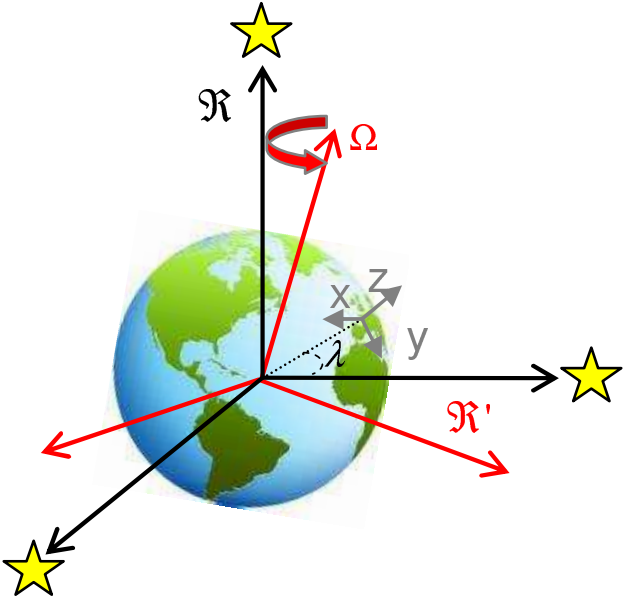}
\caption{Non rotating inertial geocentric reference frame $\mathcal{R}$ (i-frame) defined by the fixed stars having its origin at the center of the Earth and  Earth reference frame $\mathcal{R'}$ with local coordinate system $(x,y,z)$ where the $z$-axis is chosen to point away from the Earth center, with rotation rate components $\vec{\Omega}=(0,\Omega \cos\lambda,\Omega \sin \lambda)$.}
\label{fig:figure3}
\end{center}
\end{figure}

In this section we use our formulation to derive the phase shift when the atom is submitted to simultaneous time independent gravito-inertial fields like gravity, gravity gradient and rotation. We  consider the same two  cases  (A and B) treated in the previous work of \cite{Antoine2003}.\newline
In case A, the AI is fixed to the Earth frame $\mathcal{R'}$, rotating with rotation rate $\vec{\Omega}$ with respect to to the inertial reference frame (i-frame), $\mathcal{R}$ (see Fig.~\ref{fig:figure3}). In case B,  the AI is fixed to a rotating  platform of rotation rate $\vec{\Omega'}=\vec{\Omega}$ for convenience. \newline
To give the phase shift for each case (A and B), we calculate the $\vec{\mathcal{F}}_k=\left(\frac{d^{k}\vec{R}}{dt^{k}}\right)_{t=t_1}$ functions of equation ($\ref{eq:dephasage2}$). For $k=0,1$ these functions simply correspond to initial atomic position $\vec{R}_1$, and velocity $\vec{v}_1$ respectively. To calculate higher order derivatives ($k\geq 2$) we use the classical equations of atomic motion given by: 
\begin{eqnarray}
\begin{split}
\vec{g}(\vec{R})&= \underbrace{\left(\frac{d^{2}\vec{R}}{dt^{2}}\right)_{\mathcal{R'}}}_{(a)} 
+\overbrace{2\vec{\Omega}\times \left(\frac{d\vec{R}}{dt}\right)_{\mathcal{R'}}}^{(b)}\\
&+ \underbrace{\vec{\Omega}\times\left(\vec{\Omega}\times\vec{R}\right)
+\frac{d\vec{\Omega}}{dt}\times \vec{R}}_{(c)}
\label{eq:compoacc}
\end{split}
\end{eqnarray}
with $(a)$ the relative acceleration, $(b)$ the Coriolis acceleration and $(c)$ the centrifugal acceleration and where we have introduce gravity at position $\vec{R}$ as:
\begin{equation}
\vec{g}(\vec{R})=\vec{g}(\vec{R}_1)+ \vec{\vec{\Gamma}} .\left(\vec{R}-\vec{R}_{1}\right)
\label{eq:g}
\end{equation}
where $\vec{\vec{\Gamma}}$ is the gravity gradient tensor with its origin taken at position $\vec{R}_1 =\vec{R}(t=t_1)$.\newline
In order to calculate $\vec{\mathcal{F}}_k=\left(\frac{d^{k}\vec{R}}{dt^{k}}\right)_{t=t_1}$ ($k\geq 2$), one has to perform time derivation of  $(\ref{eq:compoacc})$. Moreover, one can notice that time derivation of $(\ref{eq:g})$  is strictly depending on case A or B. In order to provide a unified treatment we introduce  parameter $\eta$ to distinguish between both cases. 
Assuming $\eta=1$ for case A, and $\eta=0$ for case B, one obtains:
\begin{eqnarray}
\begin{split}
\frac{d}{dt}\left(\vec{g}(\vec{R}_1)+ \vec{\vec{\Gamma}}.(\overrightarrow{R}-\overrightarrow{R_{1}})\right)_{\mathcal{R}}=&\\
&\vec{\vec{\Gamma}}.\left(\frac{d\vec{R}}{dt}\right)_{\mathcal{R'}}
+\eta\vec{\Omega}\times\vec{g}\\
&+\eta\vec{\Omega}\times\left(\vec{\vec{\Gamma}}.(\vec{R}-\vec{R}_1\right)\\
&+(1-\eta)\vec{\vec{\Gamma}}.\left(\vec{\Omega}\times\vec{R}\right)
\label{eq:etafinal}
\end{split}
\end{eqnarray}
One interesting feature of equation ($\ref{eq:etafinal}$) is that it takes into account both cases $A$ and $B$ simultaneously.
Choosing $\eta=0$, or $\eta=1$ and calculating successive higher order time derivatives using ($\ref{eq:compoacc}$) and its derivatives and substituting in equation ($\ref{eq:dephasage2}$) allows to calculate the global inertial phase shift for any $N$ light-pulse  AIs. 
\subsubsection{Application}
Considering the rotating Earth frame, one can define $\left(\frac{d\vec{R}(t=t_1)}{dt}\right)_{\mathcal{R'}}=\vec v^{'}_{1}$ as the initial velocity of the atoms and acceleration
\begin{equation}
\vec{a}=\vec{g}(\vec{R}_1)-\vec{\Omega}\times\left(\vec{\Omega}\times\vec{R}_1\right)
\label{eq:defacc}
\end{equation}
as the sum of gravity and centrifugal accelerations. One can now calculate  the function $\vec{\mathcal{F}}_k\left(\vec{R}_1,\vec{v}_1',\vec{a},\vec{\Omega},\vec{\vec{\Gamma}}\right)$ up to order $k$.\newline
Calculating the inertial phase shift up to terms proportional to  $T^{2}$ ($k=2$) leads to :
\begin{eqnarray}
\begin{split}
\Delta \Phi_{\mathrm{inertial}}&=\frac{\Delta G_0}{0!}\vec{k}_1 \cdot \vec{R}_1 + \frac{\Delta G_1}{1!}\vec{k}_1 \cdot \vec{v}^{'}_1 T \\
&+\frac{\Delta G_2}{2!}\vec{k}_1 \cdot \left( \vec{a} -
2\vec{\Omega}\times \vec{v}^{'}_1  \right)T^2+ \text{O}(T^3)
\label{eq:PhaseGlobale}
\end{split}
\end{eqnarray}
Equation ($\ref{eq:PhaseGlobale}$) exhibits the total inertial phase shift to the second order in the time seperation beetween pulses. In this case, one can see that  $\vec{\mathcal{F}}_2$ is only a function of constant acceleration $\vec{a}$, initial velocity $\vec{v}'_1 $ and constant rotation $\vec{\Omega}$.\newline
In Table \ref{tab:tab1} we extend the phase shift calculation up to terms proportional to $T^{4}$ for whatever the case is (A or B).  In this case, cross-terms between gravity gradient $\vec{\vec{\Gamma}}$, acceleration $\vec{a}$ and rotation $\vec{\Omega}$ appear in the phase shift expression.\newline

\begin{table*}[h]
\centering
\caption{\bf Phase shift terms up to the order $k=4$.}
\begin{tabular}{lc}
\hline
& \large{\textbf{CASE}
(\textbf{A}: $\eta = 1$ and \textbf{B}: $\eta = 0$) }\\
                                                             & Rotation
$\vec{\Omega}$; gravity gradient $\vec{\vec{\Gamma}}$\\
& Acceleration $\vec{a}$ ; initial position $\vec{R}_1$; initial velocity $\vec{v^{\prime}_1}$\\
\hline
$\Delta\Phi_t$        & $\Delta G_0 . \phi_1 + \Delta G_1 \omega_1 T + \Delta
\omega\Delta G_2 \frac{T}{2}$   \\
\hline
 $T^0$           & $\frac{\Delta G_0}{0!}\vec{k}_1 \cdot \vec{R}_1$\\
\hline
 $T^1$           & $\frac{\Delta G_1}{1!}\vec{k}_1 \cdot \vec{v}^{'}_1
$\\
\hline
$T^2$           & $\frac{\Delta G_2}{2!}\vec{k}_1 \cdot \left( \vec{a} -
2\vec{\Omega}\times \vec{v}^{'}_1  \right)$\\
\hline
 $T^3$           & $\frac{\Delta G_3}{3!}\vec{k}_1 \cdot
\left(\begin{array}{c}
(\eta - 3)\cdot\vec{\Omega}\times \vec{a} +   \\  \vec{\vec{\Gamma}}\cdot
\vec{v}^{\prime}_1 + 3\cdot\vec{\Omega}\times\left(
\vec{\Omega}\times\vec{v}^{\prime}_1 \right)  \\ + (\eta -
1)\cdot\vec{\Omega}\times\left( \vec{\Omega}\times \left(
\vec{\Omega}\times \vec{R}_1 \right) \right) \\ -(\eta -
1)\vec{\vec{\Gamma}}\cdot\left( \vec{\Omega}\times \vec{R}_1 \right)
\end{array}\right)$\\
\hline
  $T^4$           & $\frac{\Delta G_4}{4!}\vec{k}_1 \cdot
\left(\begin{array}{c}
\vec{\vec{\Gamma}}\cdot \vec{a} + 3(2-\eta)\cdot\vec{\Omega}\times(\vec{\Omega}\times
\vec{a})   \\  -4 \cdot\vec{\Omega}\times \left( \vec{\Omega}\times \left(
\vec{\Omega} \times \vec{v}^{\prime}_1 \right)\right)  \\ -2\eta\vec{\vec{\Gamma}}\cdot(\vec{\Omega} \times \vec{v}^{\prime}_1)+2(\eta-2)\vec{\Omega}\times(\vec{\vec{\Gamma}}\cdot \vec{v}^{'}_1)\\ + (1 -
\eta)\cdot\vec{\vec{\Gamma}}\cdot\left( \vec{\Omega}\times \left(
\vec{\Omega}\times \vec{R}_1 \right) \right) \\ -4(1 -
\eta)\cdot\vec{\Omega}\times \left( \vec{\vec{\Gamma}}\cdot\left(
\vec{\Omega}\times \vec{R}_1 \right)\right)\\ + 3(1 -
\eta)\cdot\vec{\Omega}\times \left( \vec{\Omega}\times\left(
\vec{\Omega}\times \left(\vec{\Omega}\times\vec{R}_1 \right)\right)\right)
\end{array}\right).$\\ \hline
\end{tabular}
\label{tab:tab1}
\end{table*}
One can notice from Table~\ref{tab:tab1} that the coupling between the gravity gradient and rotation is case dependent. When the AI is fixed to a mobile platform, there is a coupling between gravity gradient and rotation whereas this coupling vanishes in case A, when the sensor is fixed to the Earth frame.\newline
Moreover, Table \ref{tab:tab1} highlights the benefit of our formulation where the phase shift terms up to $T^{k}$  dependences appear as a simple product between the inertial fields and a specific $\Delta G_k$ coefficient.\newline
The physical meaning of $\Delta G_k$ coefficients will be given in section \ref{sec:analysis} for well-known AIs.
Moreover, one would like to emphasize that this formulation may also adress the case of an optical corner cube gravimeter such as the FG-5 \cite{Niebauer1995}, by simply  nulling the laser chirp $\Delta\omega$ in Table~\ref{tab:tab1} and considering a 2 pulse interferometer with $\Delta \vec{\epsilon} =(0,1)$.
\section{Analysis and link with well known AI\small{s}}
\label{sec:analysis}
In this section we give a physical interpretation of the leading terms  $\Delta{G}_{k}$ of the phase shift which are directly related to the two-photon light pulse sequence of the interferometer through equation (\ref{eq:deltaG}).
\subsection{Impact of $\Delta G_k$ coefficients and AI symmetry}
Assuming no inertial effects and focusing on the first $\Delta G_{0}$ $(k=0)$ term of the phase shift, one can show that this term gives an information on the interferometer's closure in momentum space. This information is related to the space-time geometry of the atom interferometer.\newline
Searching for the first $N$-light-pulse interferometer closed in momentum space (i.e. $\Delta G_0=0$) with $\Delta \epsilon_1=1$, one finds out that $N\geq 2$.
For $N=2$, the simplest  atom interferometer closed in momentum space is depicted in (Fig.~\ref{fig:figure3}.(a). It is the  atomic clock configuration, also called Ramsey-Raman interferometer consisting of two pulses  $(\pi/2,t_1=0)-(\pi/2,t_2=T)$ separated by a free precession time $T=t_2-t_1$. One can see that this interferometer is time-antisymmetric, meaning that $\Delta \epsilon_1=-\Delta\epsilon_2$. 
This closure in momentum space can be expressed as:
\begin{equation}
 \Delta\epsilon_i=-\Delta\epsilon_{N+1-i}
\end{equation}
leading to $\Delta G_0=0$.\newline
The $\Delta G_{1}$ $(k=1)$ coefficient of the phase shift is related to the closure in space-time of the atom interferometer. In the Ramsey-Raman interferometer ($\Delta G_1=1$), thus the total phase shift is dependent on the initial atomic velocity and on the time dependent phase shift term. 
In order to build interferometers both closed in momentum and position space, one needs to look for $N\geq 3$ pulse interferometers.
Considering a three-pulse interferometer one has to solve for:

\begin{eqnarray}
\left\{
\begin{split}
\Delta G_0=&\Delta \epsilon_1+\Delta \epsilon_2+\Delta \epsilon_3=0\\
\Delta G_1=&\Delta \epsilon_2+ 2\Delta \epsilon_3=0  
\label{eq:3pulses}
\end{split}
\right.
\end{eqnarray}

For $\Delta \epsilon_1=1$ one finds the well-known single-loop, Mach-Zehnder configuration \cite{Leveque2009,Debs2011} consisting in the pulse sequence $(\pi/2,t_1=0)-(\pi,t_2=T)-(\pi/2,t_3=2T)$. The interferometer is depicted in (Fig.~\ref{fig:figure3}.b)). Considering two-photon Raman transitions, the first $\pi/2$ pulse puts the atom in a coherent superposition of ground and excited states and acts as a beam-splitter by transferring $\hbar k$ momentum to the wave packet making the transition to the excited state. Then, the two wave packets propagate freely during time $T$ drifting apart with relative momentum $\hbar k$. A mirror pulse is applied at time $T$ interchanging the ground and excited states and reversing their relative momenta. Finally, the two wave packets drift back towards each other during time $T$ and a final beam splitter pulse is applied at time $2T$  recombining the two wave packets and interfering them. The interferometer's closure in position space can be related to the time-symmetry through:
\begin{equation}
\Delta\epsilon_i=\Delta\epsilon_{N+1-i}
\end{equation}
 
\begin{figure}[H]
\begin{center}
\includegraphics[width=\linewidth]{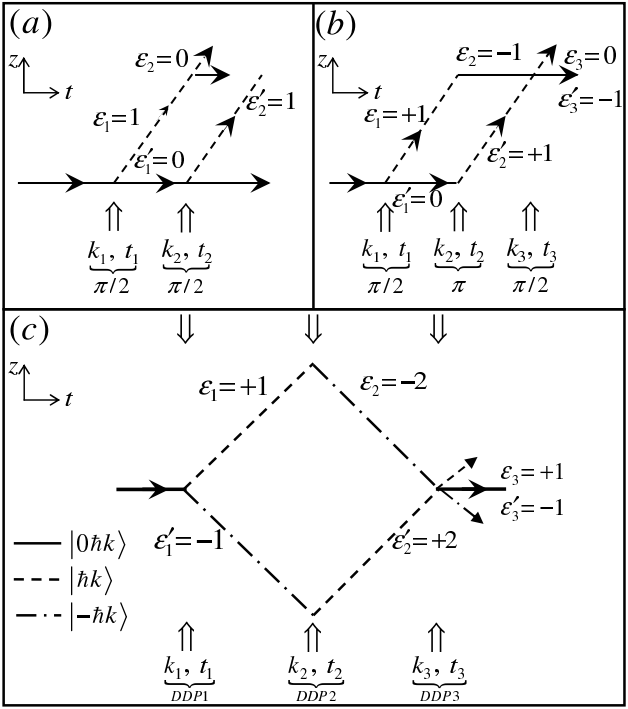}
\caption{Space-time recoil diagrams in the absence of gravity: (a) Ramsey interferometer consisting of two $\pi/2$ pulses separated by a free precession time $T=t_2 -t_1$. The interferometer is closed in momentum space $\Delta G_0=0$; (b) The Mach-Zehnder interferometer consisting of pulse sequence $(\pi/2,t_1=0)-(\pi,t_2=T)-(\pi/2,t_3=2T)$. This single-loop interferometer is closed in both momentum and position space. Its sensitivity to acceleration is proportional to the space time area enclosed by the loop. (c) Double-diffraction atom interferometer: two pairs of Raman beams with effective wave vectors $\pm k_i$ are shined simultaneously on the atoms at three moments in times seperated by time $T$. DDP1 and DDP3 are double diffraction pulses  which act as splitters whereas DDP2 acts as a mirror pulse transferring momentum $\Delta p=\mp 2\hbar k_i$ to $\pm \hbar k_i$ wave packets. The interferometer is totally symmetric and the scale factor is improved by a factor of 2. In cases (b),(c) the phase shift does not depend on the initial velocity of the atoms.}
\label{fig:figure4}
\end{center}
\end{figure}
\begin{figure}[H]
\begin{center}
\includegraphics[width=\linewidth]{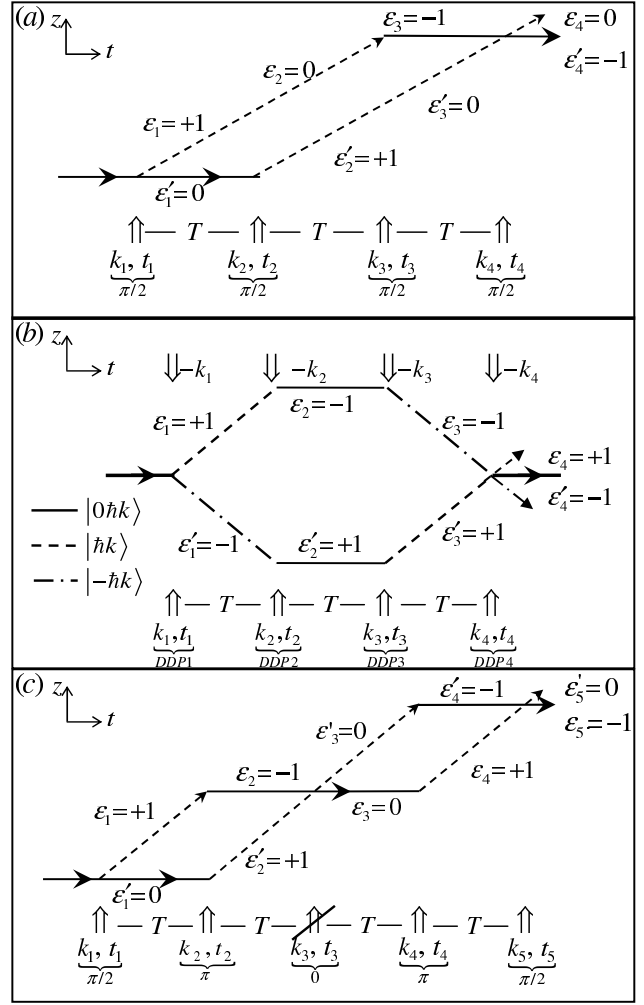}
\caption{Space time recoil diagrams in the absence of gravity: (a) Symmetric four-pulse Ramsey-Bord\'e interferometer. (b) Ramsey-Bord\'e interferometer in a double-diffraction scheme ($M=1$).(c) Five-pulse double-loop interferometer closed in both momentum and position space. In this configuration, $\Delta G_2=0$ making the interferometer insensitive to homogeneous acceleration.}
\label{fig:figure5}
\end{center}
\end{figure}

Solving the set of equations (\ref{eq:3pulses}) and	assuming $(\max|\Delta\epsilon_1|=2M, M\geq 1)$,  any three-pulse AI closed in momentum and position space obeys:
\begin{equation}
\overrightarrow{\Delta \epsilon}=M\left(\begin{array}{c} 2\\-4\\2\end{array}\right)\,\,\,
\end{equation}
For $M=1$, one finds the double-diffraction scheme using for example two-photon Raman transitions \cite{Leveque2009} and depicted in Fig.~\ref{fig:figure4}.$c)$. In this configuration two pairs of Raman beams simultaneously couple the initial ground state with zero momentum $\left |g, 0\hbar k\right >$ to the excited state with two symmetric momentum states, $\left|e,\pm\hbar k \right>$. The interferometer is time and space symmetric as $\epsilon_i=-\epsilon'_{i}$. This spatial symmetry can be expressed for any N-pulse AI as:
\begin{equation} 
\Delta\epsilon_i=2\epsilon_i
\end{equation}
The separation between the atomic wave packets leaving the first beam splitter is increased by a factor of two, leading to a difference of momenta between the two arms of $\Delta p=2\hbar k$, hence increasing the space-time area of the apparatus by a factor of two. One can see that for these two configurations closed in position space, the phase shift does not depend on the initial atomic velocity to first order (i.e. when one neglects inertial forces).\newline
When looking at four-pulse AIs, assuming $\Delta G_1=\Delta G_0=0$ one  finds the Ramsey-Bord\'e interferometer \cite{Borde1989,Muller2008,Charriere2012,Andia2013} and the double-diffraction Ramsey-Bord\'e interferometer ($M=1$) depicted in (Fig.~\ref{fig:figure5}.a,b).
Finally, cancellation of $\Delta G_{0,1,2}$ coefficients can be obtained with a five-pulse AI geometry where the middle pulse is not shined. This interferometer which is depicted in (Fig.~\ref{fig:figure5}.c), is the well known double-loop interferometer consisting of the pulse sequence $(\pi/2,t_1=0)-(\pi,t_2=T)-(\mathrm{No}\,\mathrm{pulse},t_3=2T)-(\pi,t_4=3T)-(\pi/2,t_5=4T)$. This double-loop interferometer is well suited when one wants for example to build a sensor insensitive to homogeneous acceleration for rotation or direct gravity-gradient measurements \cite{McGuirk2002}.
We give in Table~\ref{tab:tab2} the absolute values of the phase coefficient $\frac{\Delta G_k}{k!}$ for the interferometers described above.
\begin{table*}[h]
\centering
\caption{\bf Examples of phase shift coefficient  values $\frac{\Delta G_k}{k!}$ for usual atomic interferometers.}
\begin{tabular}{lccccc}
\hline
\textbf{AI}& \textbf{Clock}& \textbf{Mach-Zehnder}&\textbf{Double-Diff}&\textbf{Ramsey-Bord\'e}& \textbf{Double-loop}
\\                           
\hline
$\Delta\vec{\epsilon}$& $\left(\begin{array}{c} 
1\\-1\\0\\0\end{array}\right)$&$\left(\begin{array}{c} 
1\\-2\\1\\0\end{array}\right) $&$\left(\begin{array}{c}2\\-4\\2\\0 \end{array}\right)$&$\left(\begin{array}{c}1\\-1\\-1\\1 \end{array}\right)$&$\left(\begin{array}{c}-1\\2\\0\\-2\\1 \end{array}\right)$\\
\hline
$\frac{\Delta G_0}{0!}$&0 &0&0&0&0\\
\hline
$\frac{\Delta G_1}{1!}$&1&0&0&0&0\\
\hline
$\frac{\Delta G_2}{2!}$&$\frac{1}{2}$&1&2&2&0\\
\hline
$\frac{\Delta G_3}{3!}$&$\frac{1}{6}$&1&2&3&2\\
\hline
$\frac{\Delta G_4}{4!}$&$\frac{1}{24}$&$\frac{7}{12}$&$\frac{7}{6}$&$\frac{8}{3}$&4\\
\hline
$\frac{\Delta G_k}{k!}$&$\frac{1}{k!}$&$\frac{2^k-2}{k!}$&$\frac{2(2^k-2)}{k!}$&$\frac{3^k-2^k-1}{k!}$&$\frac{4^k-2\times 3^k +2}{k!}$\\
\hline
\end{tabular}
\label{tab:tab2}
\end{table*}

Considering the simple expression of $\Delta G$ given in equation (\ref{eq:vandermonde}), one could think of finding any interferometer configurations (i.e. any sets of vector  $\overrightarrow{\Delta \epsilon_i}$) for which the interferometer could be selectively sensitive to  gravity, acceleration, gravity gradient or rotation by fixing to zero the value of a particular $\Delta G$ coefficient and solving:  
\begin{equation}
\Delta \epsilon=V^{-1}\Delta G
\label{eq:vandermonde}
\end{equation}   
where $V^{-1}$ is the inverse Vandermonde Matrix. We will give practical examples of some N-light-pulse interferometers corresponding to the resolution of equation $(\ref{eq:vandermonde})$, in section \ref{sec:originalai}.

\subsection{Arbitrary temporal pulse-sequence}
\label{sec:arbitrarypulse}
\begin{figure}[H]
\begin{center}
\includegraphics[width=\linewidth]{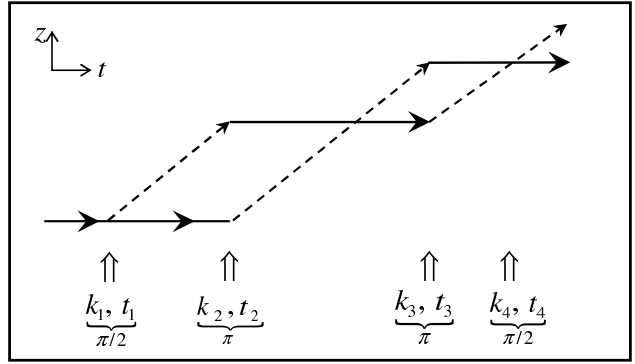}
\caption{Space-time diagram of a double loop interferometer with non identical loops. Assuming $t_1=0$, and total interrogation time $t_4=1$, in order to eliminate all $t^{3}$ terms (i.e. $\Delta G_3=0$) to the phase shift, one finds $t_{2,3}=\frac{\sqrt{5}\mp 1}{4}$. Solid and dash lines correspond to momentum states separated by $\hbar k$.}
\label{fig:figure6}
\end{center}
\end{figure}

Our approach remains consistent when considering non-equally spaced in time laser pulses. In this general case, equation ($\ref{eq:deltaG}$) has to be re-written including the instant $t_i$ of the $i$-th laser pulse leading to: 
\begin{eqnarray}
\begin{split}
\Delta G^{(t)}_k&=\sum_{i=1}^{N}\Delta \epsilon_i t_{i}^{k}\\
&=V^{(t_1,t_2,t_3,\cdots,t_N)}\Delta \epsilon
\label{eq:deltaGbis}
\end{split}
\end{eqnarray}
Consequently, $\Delta G_k$ coefficients are now time-dependent but still related to the two-photon pulse sequence of the AI through a so-called Vandermonde matrix $V^{(t_1,t_2,...,t_i,...,t_N)}$ with time-dependent coefficients.\newline
Considering a four-pulse double-loop interferometer $\Delta\vec{\epsilon}=(1,-2,-2,-1)$, one can look for solutions in time $t_i$ of the laser-pulse sequence assuming  $\Delta G_{k=0,1}=0$.
Applying equation (\ref{eq:vandermonde}) with respect to $V^{(t_1,t_2,t_3,t_4)}$ leads to:
\begin{equation}
\Delta \epsilon=(V^{(t_1,t_2,t_3,t_4)})^{-1}\Delta G
\label{eq:vandermondetemps}
\end{equation}
The detailed calculations are given in Appendix $B$ considering two cases and assuming $t_1=0$ and $t_4=1$ in dimensionless unit.\newline
In the first case we assumed $\Delta G_2=0$. If one refers to Table~\ref{tab:tab1}, in this case, the phase dependent terms which scale quadratically with time $t$ are eliminated. This is interesting when one wants to measure gravity-gradients or rotations. It then comes out that the pulse-sequence is $\lbrace{t_2=\frac{1}{4}t_4;t_3=\frac{3}{4}t_4\rbrace}$ leading to $\Delta G_3=\frac{3}{16}t_{4}^{3}$.
This pulse-sequence corresponds exactly to the case of the identical double-loop interferometer found in  section 3.A.  when one was not varying the inter-pulse time  (Fig~\ref{fig:figure5}.c).
In the second case we assumed $\Delta G_3=0$, hence eliminating all phase terms scaling as $t^{3}$. This case is interesting when one wants to increase the accuracy to acceleration measurement by eliminating $T^3$ corrections to the phase. We represent in Fig.~\ref{fig:figure6} the pulse sequence $\lbrace{t_2=\frac{\sqrt{5}-1}{4};t_3=\frac{\sqrt{5}+1}{4}\rbrace}$ corresponding to a non-identical double-loop interferometer leading to $\Delta G_2=\left(1-\frac{\sqrt{5}}{2}\right)t_{4}^{2}$ as was found in the previous work of \cite{Dubetsky2006} where phase shift calculations were done using density matrix formalism.

\section{Effect of wave vector change and photon recoil on the phase shift}
\label{sec:photonrecoil}
\subsection{Wave vector change}
Considering the wave vector change then one has to add a correction to equation (\ref{eq:dephasage}) leading to:
\begin{eqnarray}
\begin{split}
&\Delta \Phi_{\mathrm{inertial}} \simeq\\
&\sum_{k=0}\frac{\Delta G_k+\frac{\Delta k}{k_1}\Delta G_{k+1}}{k!}\cdot T^k
&\vec{k}_{1}\cdot \vec{\mathcal{F}}_k\left(\vec{R}_1,\vec{v}_1,\vec{a},\vec{\Omega},\vec{\vec{\Gamma}}\right)\\
\end{split}
\label{eq:dephasagedeltak}
\end{eqnarray}
where one can see that the wave vector change contribution to the phase shift at the order $k$ in the time separation between pulses, is simply obtained from the calculation of $\Delta G_{k+1}$ coefficient.
\subsection{Recoil phase shift}
\begin{figure}[H]
\centering
\includegraphics[width=\linewidth]{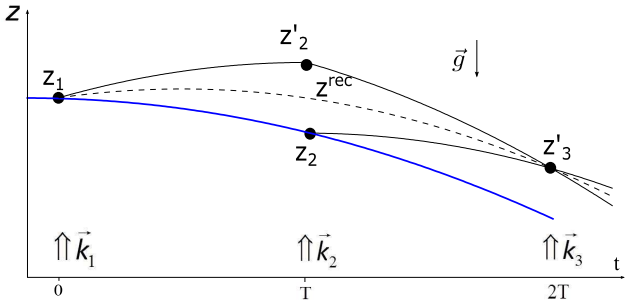}
\caption{Space-time diagram showing the Mach-Zehnder interferometer paths in the presence of gravity acceleration. Solid lines: upper and lower interferometer paths ($z'_i$ and $z_i$). Dash line: classical mid-point positions of the atoms taking into account the recoil effect; Solid line (in blue): parabola in the absence of photon recoil. The trajectories are plotted assuming $k_1=k_2=k_3$.}
\label{fig:figure7}
\end{figure}

The two-photon transitions contribute to transfer two photon momenta to the atoms. Thus, the inertial phase shift is modified and one needs to account for $z_i\neq z'_i$.
Assuming no rotation and no inhomogeneous acceleration (i.e. gravity gradient), one can express simply the position of the atoms under free-fall at time $t_i$ considering the recoil velocity term $\hbar k/m$ for an atom of atomic mass $m$:

\begin{multline}
z_i=z_1+v_1(i-1)T +\underbrace{\frac{\hbar}{m}T\sum_{j=1}^{i-1}(i-j)k_j\epsilon_j}_{(a)}\\
+\frac{1}{2}g(i-1)^2T^2 +\text{O}(T^3),\\
z'_i=z_1+v_1(i-1)T+\underbrace{\frac{\hbar}{m}T\sum_{j=1}^{i-1}(i-j)k_j\epsilon'_j}_{(b)}\\
+\frac{1}{2}g(i-1)^2T^2+\text{O}(T^3),
\end{multline}

where $(a)$ and $(b)$ are the recoil dependent terms associated to the two-photon transition for the upper and lower path respectively.
Hence, calculating the classical mid-point position of the atoms one finds:
\begin{eqnarray}
\begin{split}
z_{i}^{rec}=\frac{z_i+z'_i}{2}&=z_1+v_1(i-1)T+\frac{1}{2}g(i-1)^2T^2\\
&+\underbrace{\frac{\hbar}{2m}T\sum_{j=1}^{i-1}(i-j)k_j(\epsilon_j+\epsilon_j')}_{(c)}
\end{split}
\label{eq:recoilterm}
\end{eqnarray}
where $(c)$ represents the recoil effect contribution to the inertial phase shift.
In Fig~\ref{fig:figure7} we represent the space-time recoil  interferometer paths and the classical mid-point position of the atoms in a Mach-Zehnder geometry with pulse-sequence $(\pi/2 ,t_1=0)-(\pi,t_2=T)-(\pi/2,t_3=2T)$.

Considering the two-photon momenta transfer the inertial phase shift cannot be approximated anymore by equation ($\ref{eq:dephasageaprox}$) and one has to account for the recoil effect contribution of ($\ref{eq:recoilterm}$).
Hence, the inertial phase shift is:
\begin{eqnarray}
\begin{split}
\Delta\Phi_{\mathrm{inertial}}&=\sum_{i=1}^{N}k_{i}\Delta\epsilon_{i}z_{i}^{rec} \\
&=\sum_{i=1}^{N}k_{i}\Delta\epsilon_{i}z_i +\Delta\Phi_{rec}
\end{split}
\end{eqnarray}
where the first term is the inertial phase shift given by  equation~($\ref{eq:dephasagedeltak}$)  in the absence of photon recoil, and the second term is the recoil phase shift contribution that can be expressed in the time separation between pulses as:
\begin{equation}
\Delta \Phi_{\mathrm{rec}}=\frac{\hbar}{2m}k^2T\Delta A_1+\frac{\hbar}{2m}k\Delta k T\Delta B_1 +\frac{\hbar}{2m}\Delta k^2 T \Delta C_1 + \text{O}(T^2)
\label{eq:dephasagerecul}
\end{equation}
where we introduce :
\begin{eqnarray}
\begin{split}
\Delta A_{k}&=\sum_{i=1}^{N}(\epsilon_i-\epsilon'_i)\times \\
&\left(\sum_{j=1}^{i-1}
(i-j)^{k}(\epsilon_j+\epsilon'_j)\right)
\end{split}
\end{eqnarray}

\begin{eqnarray}
\begin{split}
\Delta B_{k}&=\sum_{i=1}^{N}(\epsilon_i-\epsilon'_i)\times \\
&\left(\sum_{j=1}^{i-1}
(i-1+j-1)(i-j)^{k}(\epsilon_j+\epsilon'_j)\right)
\end{split}
\end{eqnarray}

\begin{eqnarray}
\begin{split}
\Delta C_{k}&=\sum_{i=1}^{N}(\epsilon_i-\epsilon'_i)\times \\
&\left(\sum_{j=1}^{i-1}(j-1)(i-j)^{k}(i-1)(\epsilon_j+\epsilon'_j)\right)
\end{split}
\end{eqnarray}
for $k\geq 0$.\newline
The first term of equation (\ref{eq:dephasagerecul}) is the well known recoil shift associated to the kinetic energy of an atom absorbing a photon of momentum $\hbar k$, and  the two last terms depend on the wave vector change $\Delta k$ to the first and second order respectively. These terms may account for corrections to gravity measurement when one considers the variation of the laser frequencies to compensate the Doppler shift of freely falling atoms \cite{Wolf1999}.\newline
One can generalize the recoil phase shift to higher orders in the time separation between pulses by simply writing:
\begin{eqnarray}
\begin{split}
\Delta\Phi_{\mathrm{rec}}=\\
&\sum_k\frac{\Delta A_k+\frac{\Delta k}{k_1}\Delta B_k+(\frac{\Delta k}{k_1})^2\Delta C_k}{k!}\cdot T^k\\
&\times \vec{k}_1\cdot\vec{\mathcal{F}}_k\left(\vec{0},\vec{v}_r=\frac{\hbar \vec{k}_1}{2m},\vec{0},\vec{\Omega},\vec{\vec{\Gamma}}\right)
\end{split}
\end{eqnarray}
where the $\vec{\mathcal{F}}_k$ function takes into account the recoil velocity increment $\vec{v}_r$ in the presence of laser interaction. Thus the recoil velocity is coupled to inertial forces.
The photon recoil phase shift contribution is given in Table ~\ref{tab:tab3} assuming $\Delta k=0$ for simplicity.
Nevertheless, $\Delta k\neq 0$ phase shift terms can be obtained by calculating $\Delta B_k$ and $\Delta C_k$ coefficients.
\begin{table*}[htbp]
\centering
\caption{\bf Recoil phase shift terms up to the order $T^4$ ($k=4$) assuming $\Delta k=0$.}
\begin{tabular}{lc}
\hline
\bf order k &\bf Phase Shift  \\
\hline
 $k=0$           & 0\\
\hline
 $k=1$           & $\frac{\Delta A_1}{1!}\vec{k}_1\cdot\frac{\hbar \vec{k}_1}{2m}\cdot T$\\
\hline
$k=2$           & 0\\
\hline
 $k=3$           & $\frac{\Delta A_3}{3!}\vec{k}_1 \cdot
\left(\begin{array}{c}
3\vec{\Omega}\times \left(\vec{\Omega} \times \frac{\hbar \vec{k}_1}{2m}\right)  \\ 
+\vec{\vec{\Gamma}}\cdot\left( \frac{\hbar \vec{k}_1}{2m}\right)
\end{array}\right) T^3 $\\
\hline
 $k=4$           &-2 $\frac{\Delta A_4}{4!}\vec{k}_1 \cdot
\left(\begin{array}{c}
(2-\eta)\cdot \vec{\Omega}\times \left(\vec{\vec{\Gamma}}\cdot\frac{\hbar \vec{k}_1}{2m}\right) \\  
\eta\vec{\vec{\Gamma}}\cdot\left(\vec{\Omega} \times \frac{\hbar \vec{k}_1}{2m}\right)
\end{array}\right) T^4 $\\ \hline
\end{tabular}
\label{tab:tab3}
\end{table*}

We give in Table~\ref{tab:tab4} the $\Delta A_k/k!$ terms for usual interferometers.

\begin{table*}[htbp]
\centering
\caption{\bf Examples of recoil phase shift coefficient  values $\frac{\Delta A_k}{k!}$  for usual atomic interferometers.}
\begin{tabular}{lccccc}
\hline
\textbf{AI}& \textbf{Clock}& \textbf{Mach-Zehnder}&\textbf{Double diffraction}&\textbf{Ramsey-Bord\'e}&\textbf{Ramsey-Bord\'e-Chu}\\                           
\hline
$\Sigma \vec{\epsilon}$& $\left(\begin{array}{c} 
1\\1\end{array}\right)$&$\left(\begin{array}{c} 
1\\0\\-1\end{array}\right) $&$\left(\begin{array}{c} 
0\\0\\0\end{array}\right) $&$\left(\begin{array}{c}1\\1\\-1\\-1 \end{array}\right)$&$\left(\begin{array}{c}1\\-1\\-1\\1 \end{array}\right)$\\
\hline
$\frac{\Delta A_1}{1!}$&1&0&0&0&-2\\
\hline
$\frac{\Delta A_3}{3!}$&$\frac{1}{6}$&1&0&$4$&$\frac{5}{3}$\\
\hline
$\frac{\Delta A_4}{4!}$&$\frac{1}{24}$&$\frac{7}{12}$&0&$\frac{39}{12}$&$2$\\
\hline
\end{tabular}
\label{tab:tab4}
\end{table*}

One can note that contrary to the Ramsey-Bord\'e-Chu AI \cite{Weiss1993} consisting of two pairs of $\pi/2$ pulses where the second pair of $\pi/2$ pulses propagate in the opposite direction to the first pair, the Ramsey-Bord\'e interferometer (Fig~\ref{fig:figure4}(a)) used to determine the fine structure constant through a measurement of the atom recoil velocity in \cite{Cadoret2008,Bouchendira2011}  is  not sensitive to the recoil shift at the first order in time $T$ between pulses (ie $\Delta A_1=0$) although recoil velocity dependent terms appear as corrections to the phase in power of $k\geq 3$ in time separation between pulses. This main difference comes from the aditional kinetic energy term acquired by the atoms in one of the arm of the Ramsey-Bord\'e-Chu interferometer, inducing a photon recoil dependence of the phase. However, in order to measure the atomic recoil in the Ramsey-Bord\'e interferometer, one has to add photon recoil to the atoms in between the two sets of beam splitters. This can be done by implementing a sequence of Bloch oscillations as in \cite{Cadoret2008,Bouchendira2011}, thus leading to a significant sensitivity increase in the recoil measurement.\newline
Moreover, recoil phase shift terms presented in Table~\ref{tab:tab3} are responsible for not perfectly closed AIs due to coupling between the atomic recoil with rotations and gravity gradient forces.  Focusing on the  Mach-Zehnder AI used as a gravimeter with vertically propagating optical pulses (along the $z$-axis), this means that the classical trajectories of the upper path and the lower path will not exactly intersect at the final beam splitter. The separation distance between the wave packets at the last pulse  is  given by $\Delta z=\partial \Delta\Phi_{rec}/\partial k_1\approx \frac{\Delta A_3}{3!}\frac{\hbar k_z\Gamma_{zz}}{2m}T^3$ when one only looks at the contribution of gravity gradient force.\newline
Considering a $^{87}$Rb atom interferometer ($\lambda$=780 nm, $T=1$ s, $k_z=\frac{4\pi}{\lambda}$), one finds a separation between the wave packets of $\Delta z\simeq 18\,\mathrm{nm}$. This separation distance is not a problem  when using $\mu$K-temperature thermal cloud sources obtained from usual optical molasses. Nevertheless, this separation phase shift could become an issue for long baseline AIs (typically $T>5$ s) where BEC-based AIs are required \cite{Carraz2014}. 

\subsection{Total phase shift}
The total phase shift of an N-light-pulse AI is simply given by:
\begin{equation}
\Delta\Phi_{\mathrm{Total}}=\Delta\Phi_t+\Delta\Phi_{\mathrm{inertial}}
\end{equation}
where the total inertial phase shift takes into account the wave vector mismatch and recoil effect.\newline
One finds:
\begin{eqnarray}
\begin{split}
\Delta \Phi_{\mathrm{Total}}=&\Delta G_{0}\phi_{1}+\Delta G_{1}\omega_1 T+\Delta G_{2}\Delta\omega\frac{T}{2}\\&+\sum_{k}\frac{\Delta G_k+\frac{\Delta k}{k_1}\Delta G_{k+1}}{k!}\cdot T^k\\
&\times \vec{k}_{1}\cdot\vec{\mathcal{F}}_k\left(\vec{R}_1,\vec{v}_1,\vec{a},\vec{\Omega},\vec{\vec{\Gamma}}\right)\\
&+\frac{\Delta A_k+\frac{\Delta k}{k_1}\Delta B_k+(\frac{\Delta k}{k_1})^2\Delta C_k}{k!}\cdot T^k\cdot\\
&\times \vec{k}_1\cdot\vec{\mathcal{F}}_k\left(\vec{0},\vec{v}_r=\frac{\hbar \vec{k}_1}{2m},\vec{0},\vec{\Omega},\vec{\vec{\Gamma}}\right)
\label{eq:phasetotal}
\end{split}
\end{eqnarray}
Where the $\vec{\mathcal{F}}_k$ functions are given in Table~\ref{tab:tab1} and Table~\ref{tab:tab3}.
One can now calculate the total phase shift  of any N-light-pulse AI for any cases ($A,B$) of section~\ref{sec:calculations}.\newline

\section{Application to the Mach-Zehnder AI}
\label{sec:application}
As an application, we use our formulation to calculate the total phase shift $\Delta\Phi_{MZ}$ of the well-known Mach-Zehnder AI.
According to our formulation, the temporal symmetries of the interferometer implies $\Delta A_k=\Delta G_k;\Delta B_k=\Delta G_{k+1};\Delta C_k=0$. Considering equation ($\ref{eq:phasetotal}$) this leads to a simple expression of the phase shift in terms of $\vec{\mathcal{F}}_k$ function:
\begin{eqnarray}
\begin{split}
\Delta \Phi_{\mathrm{MZ}}=\Delta \omega T +
\sum_{k}\frac{\Delta G_k+\frac{\Delta k}{k_1}\Delta G_{k+1}}{k!}&\cdot T^k\\
\times \vec{\mathcal{F}}_k\left(\vec{R}_1,\vec{v}_1+\frac{\hbar \vec{k}_1}{2m},\vec{a},\vec{\Omega},\vec{\vec{\Gamma}}\right)
\end{split}
\end{eqnarray}
where $\Delta \omega T$ is the time dependent laser phase shift which depends on the laser frequency chirp applied to the lasers.\newline
As an application, we  evaluate phase terms with parameters corresponding to our transportable cold $^{87}$Rb atom gravimeter based on a $(\pi/2-\pi-\pi/2)$  two-photon Raman pulse sequence, described in \cite{Bidel2013}. The time between pulses is $T=48$ ms with initial velocity components $\vec{v}_{1}'=(v_{x,y}=v_\perp,v_z)$, where $v_\perp \simeq 1,2$ cm$/$s correspond to the transverse velocity of the atoms and  $v_z\simeq\,15$ cm/s corresponding to the vertical velocity at the first light-pulse. All phase terms are calculated considering the sensor fixed in the local Earth frame defined in section~\ref{sec:calculations}. Results are given in Table~\ref{tab:tabgravi}.

\begin{table*}[htbp]
\centering
\caption{\bf Contribution to the phase for atomic gravimeter of reference \cite{Bidel2013} in the rotating Earth frame. For numerical applications : effective wave vector of the rubidium two-photon transition: $k=k_z=\frac{4\pi}{\lambda}=1.61\times 10^7$ m$^{-1}$ with $\lambda=\,780$ nm the effective wavelength of the transition, Earth rotation rate: $\Omega= 7.29\times 10^{-5}$ rad/s ; latitude : $\lambda=48,71^{\circ}$ and vertical gravity gradient: $\Gamma_{zz}=3.1\times 10^{-6}$ s$^{-2}$ (assuming a spherical symmetric Earth), with $g_z= 9.81$ m/s$^{2}$.The gravimeter phase shift $k_z gT^{2}$ is taken as a reference. }
\begin{tabular}{lcr}
  \hline
  \textbf{Phase term}& \textbf{Numeric value [rad]}& \textbf{Relative phase}  \\
  \hline
  $k_{\mathrm{z}}g T^2$ & $-3.64\times 10^{5}$ & 1 \\
  $2k_{\mathrm{z}}\Omega_y v_{\perp}T^{2}$ & $42.82\times 10^{-3}$ & $-1\times 10^{-7}$ \\
	 $k_{z}\Gamma_{zz}v_{z}T^{3}$ & $-0.8\times 10^{-3}$ & $2.2\times 10^{-9}$\\
	 $3k_{\mathrm{z}}(-\Omega_{y}^{2}v_z+\Omega_z\Omega_y v_{\perp})T^{3}$ & $-1.7\times 10^{-6}$ & $4.6\times 10^{-12}$\\
	$-k_{\mathrm{z}}\Omega_y^{2}\frac{v_{rec}}{2}T^{3}$ & $-2.42\times 10^{-8}$ & $6.65\times 10^{-14}$\\
	$k_{\mathrm{z}}\Gamma_{zz}\frac{v_{rec}}{2}T^{3}$ & $3.14\times 10^{-5}$ & $-8.6\times 10^{-11}$\\
	$\frac{7}{12}k_{\mathrm{z}}\Gamma_{zz}gT^{4}$& $-1.46\times 10^{-3}$ & $4\times 10^{-9}$\\
	$\frac{7}{12}k_{\mathrm{z}}(-4\Omega_{y}^{3}v_{\perp}-4\Omega_z^{2}\Omega_yv_{\perp})T^{4}$ & $-6.1\times 10^{-13}$ & $1.6\times 10^{-18}$\\
	$3\Delta k_{\mathrm{z}}g T^2$ & $-1.71\times 10^{-3}$ & $4.7\times 10^{-9}$ \\
	\hline
\end{tabular}
\label{tab:tabgravi}
\end{table*}

Comparison of our Mach-Zehnder AI phase shift determination can be made with previous work of
Antoine et al, Dubetsky et al and Wolf et al \cite{Antoine2003,Dubetsky2006,Wolf1999} with a simple change of variable, considering case A or B of section~\ref{sec:calculations}.As an example, in the work of Antoine et al.\cite{Antoine2003}, the wave vector change is not considered ($\Delta k=0$) and gravity is defined as:
\begin{equation}
\vec{g}_1=\vec{g}(\vec{R}_1) +\vec{\vec{\Gamma}}\cdot(\vec{R}-\vec{R}_1)
\end{equation}
where $g_1$ is gravity acceleration defined at the first light-pulse.
In previous work of \cite{Dubetsky2006} one has to consider the rotating Earth frame with initial velocity of the atoms $\vec{v}'_1=\vec{v}_1-\vec{\Omega}\vec{R}_1$ with acceleration :
\begin{equation}
\vec{g}_1=\vec{a}(\vec{R_1}) + \vec{\Omega} \times (\vec{\Omega} \times \vec{R}_1)
\end{equation}
where acceleration $\vec{a}$ is defined in equation $(\ref{eq:defacc})$.
Finally, in the work of Wolf et al.\cite{Wolf1999} who studied the effect of wave vector change ($\Delta k \neq 0$) one has to consider our formulation in case $B: \eta=0$ in an inertial frame in the absence of rotation.
In all cases we found all phase shifts to be consistent with our novel formulation.
\section{Original N-light-pulse AI sensor}
\label{sec:originalai}
Our simple formulation allows to seek for original N-light-pulse AIs  where the phase shift dependences to specific inertial effects could be cancelled, by simply searching solutions to equation ($\ref{eq:vandermonde}$) and zeroing specific $\Delta G_k$ coefficients. We give hereafter two theoritical examples of such AIs that would be dedicated to the measurement of the photon recoil.
\subsection{Photon recoil measurement AIs}
Direct and sensitive recoil frequency measurements using a four-pulse Ramsey-Bord\'e-Chu AI  can be used to determine the fine structure constant \cite{Weiss1993,Muller2009}. Considering Table~\ref{tab:tab1},Table~\ref{tab:tab3} and Table~\ref{tab:tab4} the atomic phase shift up to the power 2 in time $T$ in the rotating Earth (ie: Case A $\eta=1$) is :
\begin{equation}
\Delta\Phi_{\mathrm{Total}}=\Delta\omega T + M^2\omega_{\mathrm{r}} T+ M\vec{k}_1 (\vec{g}-2\vec{\Omega} \times \vec{v}_1')T^{2}+\text{O}(T^3)
\label{eq:chu}
\end{equation}
where $\omega_{\mathrm{r}}=\hbar \vec{k}_1^2/(2m)$ is the recoil frequency of the two photon transition of effective wave vector $k_1$, and $M$ is the Bragg diffraction order when using Bragg pulses as LMT beam splitters like in \cite{Muller2009} to enhance the sensitivity to the recoil shift by a factor $M^2$.\newline
One can see from equation $(\ref{eq:chu})$ that the phase shift is sensitive to inertial forces.\newline
In order to make  measurements only sensitive to the recoil shift, conjugate interferometer geometries are used to remove the sensitivity to local gravitational acceleration and moreover, simultaneous conjugate interferometers (SCIs) help rejecting common-mode vibrational noise \cite{Muller2009} that may affect the interferometer sensitivity. However, for example, gravity gradient effect which scales cubically with time $T$ (see Table~\ref{tab:tab1}) cannot be totally rejected as the two conjugate interferometers are separated in space.\newline
We give hereafter an illustration of how to use our simple formulation to find a theoritical N-pulse AI scheme sensitive to the recoil frequency and independent of gravity acceleration, gravity gradient and Earth rotation.

\subsubsection{Example 1: measurement independent of gravity}
\begin{figure}[H]
\begin{center}
\includegraphics[width=\linewidth]{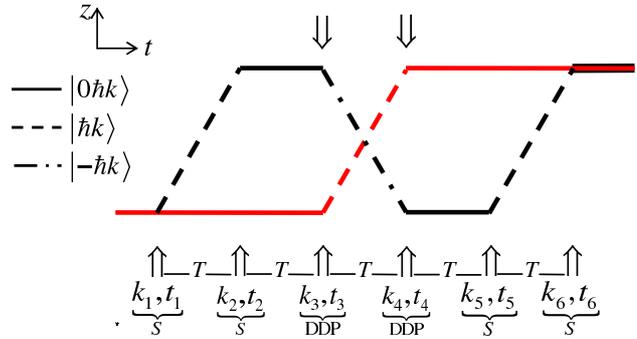}
\caption{Space-time recoil diagram of the 6-pulse atom interferometer sensitive to recoil phase shift and insensitive to gravity acceleration. Black line: (upper path) $\vec{\epsilon}=(1,-1,-1,1,1,-1)$ and Red line : (lower path) $\vec{\epsilon}^{\prime}=(0,0,1,-1,0,0)$. S: Beam-splitter pulse ; DDP:Double-Diffraction Pulse.}
\label{fig:figure8}
\end{center}
\end{figure}

One can look for an AI that would measure the photon recoil independent of the local gravity acceleration. In this case, one would have to solve equation $(\ref{eq:vandermonde})$ assuming $\Delta G_{k=0,1,2}=0$ and $\Delta A_1\neq 0$. If one considers equal time $T$ between the two-photon laser pulses, and assuming $max |\Delta\epsilon|=2$, one finds the $N=6$-light-pulse sequence of Fig~\ref{fig:figure8} where we recall that the two-photon wave vector is not adressed in terms of direction. Nevertheless, one can see that for the third and fourth pulse one needs to realize a double-diffraction scheme. The beam splitters of the interferometer are denoted $S$.
We give in Table~\ref{tab:6pulses} the phase shift coefficient values of the interferometer.

\begin{table*}[htbp]
\centering
\caption{\bf Phase shift coefficients of the 6-light-pulse atom interferometer of Fig~\ref{fig:figure8}.}
\begin{tabular}{lcr}
  \hline
  \textbf{Coefficient}& \textbf{Value}\\
  \hline\hline
  $\Delta \vec{\epsilon}$&$\left(1,-1,-2,2,1,-1\right)$  \\
\hline
  $\frac{\Delta G_0}{0!}$ & $0$ \\
	\hline
	  $\frac{\Delta G_1}{1!}$& $0$ \\
		\hline
	  $\frac{\Delta G_2}{2!}$& $0$ \\
		\hline
	 $\frac{\Delta G_3}{3!}$& $-4$ \\
	\hline
	 $\frac{\Delta G_4}{4!}$ & $10$ \\
	\hline
	 $\frac{\Delta G_5}{5!}$& $-14$ \\
	\hline
	 $\frac{\Delta A_1}{1!}$& $-2$ \\
	\hline
	$\frac{\Delta A_3}{3!}$& $\frac{-1}{3}$ \\
	\hline
\end{tabular}
\label{tab:6pulses}
\end{table*}

One can verify that the total phase shift to the first order in $T$ is:
\begin{equation}
\Delta\Phi^{\mathrm{6pulse}}_{\mathrm{Total}}=-2\frac{\hbar k_1^{2}}{2m}T=-2\omega_{\mathrm{r}}T + \text{O}(T^2)
\end{equation}
whereas all terms scaling quadratically with time $T$ are removed. However, gravity gradient and Earth rotation rate remains present within the terms scaling cubically and quartically with time $T$ (ie $\Delta G_3$ and $\Delta G_4$ coefficients).

\subsubsection{Example 2: measurement independent of gravity, gravity gradient and Earth rotation}
\begin{figure}[H]
\begin{center}
\includegraphics[width=\linewidth]{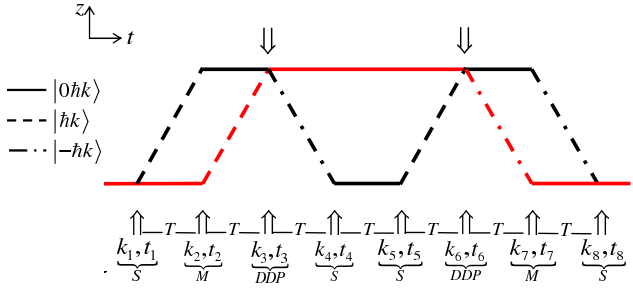}
\caption{Space-time recoil diagram of the 8-pulse atom interferometer sensitive to recoil phase shift and insensitive to Earth rotation, gravity acceleration and its gradient. Black line: (upper path) $\vec{\epsilon}=(1,-1,-1,1,1,-1,-1,1)$ and Red line : (lower path) $\vec{\epsilon}^{\prime}=(0,1,-1,0,0,-1,1,0)$. S: Beam-splitter pulse ; M: Mirror pulse ; DDP: Double-Diffraction Pulse.}
\label{fig:figure9}
\end{center}
\end{figure}

\begin{figure}[H]
\begin{center}
\includegraphics[width=\linewidth]{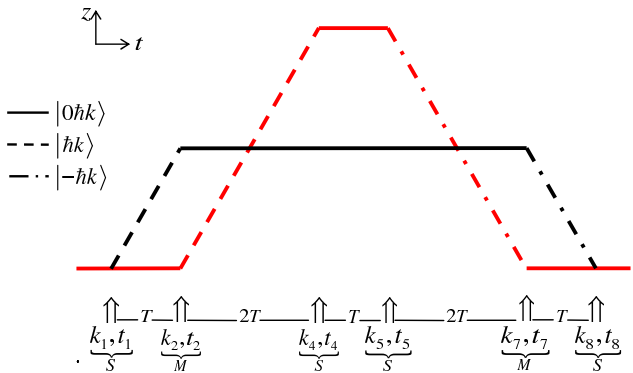}
\caption{Space-time recoil diagram of an 8-pulse atom interferometer sensitive to recoil phase shift and insensitive to Earth rotation, gravity acceleration and its gradient. Black line: (upper path) $\vec{\epsilon}=(1,-1,0,0,0,0,-1,1)$ and Red line : (lower path) $\vec{\epsilon}^{\prime}=(0,1,0,-1,-1,0,1,0)$. S: Beam-splitter pulse ; M: Mirror pulse.}
\label{fig:figure10}
\end{center}
\end{figure}

We looked for an AI configuration  insensitive to gravity gradient and Earth rotation. For this case we solved $\Delta G_{k=0,1,2,3}=0$ with $\Delta A_1 \neq0$. Considering for simplicity, optical pulses equally spaced with time $T$, and assuming $max|\Delta \epsilon|=2$, we found several 8-light pulse AI configurations. We give two of these configurations on Fig~\ref{fig:figure9} and Fig.~\ref{fig:figure10}. We give in Table~\ref{tab:8pulses} and Table~\ref{tab:8pulsesbis} the phase shift coefficient values of the two degenerate interferometer configurations. One can see that the main difference appears in the $\Delta A_3/3!$ coefficient where rotation and gravity gradient are coupled to the atomic recoil leading to a difference of almost an order of magnitude in between the two coefficients.

\begin{table*}[htbp]
\centering
\caption{\bf Phase shift coefficients of the 8-light-pulse atom interferometer  of Fig~\ref{fig:figure9}.}
\begin{tabular}{lcr}
  \hline
  \textbf{Coefficient}& \textbf{Value}\\
  \hline\hline
  $\Delta \vec{\epsilon}$&$\left(1,-2,0,1,1,0,2,1\right)$ \\
\hline
  $\frac{\Delta G_0}{0!}$ & $0$ \\
	\hline
	  $\frac{\Delta G_1}{1!}$& $0$ \\
		\hline
	  $\frac{\Delta G_2}{2!}$& $0$ \\
		\hline
	 $\frac{\Delta G_3}{3!}$& $0$ \\
	\hline
	 $\frac{\Delta G_4}{4!}$ & $6$ \\
	\hline
	 $\frac{\Delta G_5}{5!}$& $21$ \\
	\hline
	 $\frac{\Delta A_1}{1!}$& $-2$ \\
	\hline
	$\frac{\Delta A_3}{3!}$& $\frac{1}{3}$ \\
	\hline
\end{tabular}
\label{tab:8pulses}
\end{table*}

\begin{table*}[htbp]
\centering
\caption{\bf Phase shift coefficients of the 8-light-pulse atom interferometer  of Fig~\ref{fig:figure10}.}
\begin{tabular}{lcr}
  \hline
  \textbf{Coefficient}& \textbf{Value}\\
  \hline\hline
  $\Delta \vec{\epsilon}$&$\left(1,-2,0,1,1,0,-2,1\right)$  \\
\hline
  $\frac{\Delta G_0}{0!}$ & $0$ \\
	\hline
	  $\frac{\Delta G_1}{1!}$& $0$ \\
		\hline
	  $\frac{\Delta G_2}{2!}$& $0$ \\
		\hline
	 $\frac{\Delta G_3}{3!}$& $0$ \\
	\hline
	 $\frac{\Delta G_4}{4!}$ & $6$ \\
	\hline
	 $\frac{\Delta G_5}{5!}$& $21$ \\
	\hline
	 $\frac{\Delta A_1}{1!}$& $2$ \\
	\hline
	$\frac{\Delta A_3}{3!}$& $-\frac{11}{3}$ \\
	\hline
\end{tabular}
\label{tab:8pulsesbis}
\end{table*}

These theoritical AI schemes consider perfect beam splitter pulses. However, practically, the use of non perfect beam splitters induce many two-wave interferences leading to spurious phase shifts and possible contrast loss~\cite{Trebst2001}. This multiple two-wave interference effect is not treated with our theoritical framework. Nevertheless, from a practical point of view, experimental strategies to mitigate this effect could be developped such as the use of a pushing beam at resonance with the atoms to suppress parasitic paths in the interferometer~\cite{Charriere2012}.
\section{Conclusion}
\label{sec:conclusion}
In this work we have presented a novel formulation of the phase shift in multipulse atom interferometers considering two-path interferences in the presence of multiple gravito-inertial fields. We showed, considering constant acceleration and rotation and starting from an exact phase shift formula obtained in \cite{Antoine2003}, that coefficient in the leading terms of the phase shift up to $T^{k}$ dependences could be simply expressed as the product between a so-called Vandermonde matrix, and a vector characterizing the two-photon pulse sequence of the AI.\newline
We showed that this formulation could be of practical interest when one wants to use atom interferometers as a selective sensor of some specific inertial field. As an application we presented theoritical examples of original AIs  that could be dedicated to  photon recoil measurement independent of rotation, gravity, and gravity gradient. We therefore think that this formulation could benefit to the community.\newline
A noteworthy attribute of our work is that time dependent accelerations or rotations, could be easily included in our formulation, hence allowing to look for AI configurations where time dependent rotations would be minimized or suppressed as one knows their detrimental effects on atomic gradiometer performances \cite{Carraz2014}.
Finally, multiple-beam atom interferometers have demonstrated to be of interest for precision measurements of physical quantities \cite{Morinaga2001,Morinaga2003}. Therefore, a possible improvement of this work would be to consider the effect on the interferometer's phase shift of the greater number of interfering paths when dealing with  multipulse scheme atom interferometers. 
\appendix
\clearpage
\section*{Appendix A: Calculation of section 3}
\setcounter{equation}{0}

\subsection{Derivation of \textbf{section 3.B.}}
We assume a $N=4$ pulse AI with $\Delta G_{0}=\Delta G_{1}=0$.
Starting from equation (\ref{eq:vandermondetemps}) one finds:
\begin{eqnarray}
\begin{split}
\Delta\epsilon_1=\frac{[(t_1+t_2+t_3+t_4)-t_1]\Delta G_2-\Delta G_3}{(t_2-t_1)(t_3-t_1)(t_4-t_1)}\\
\Delta\epsilon_2=\frac{[(t_1+t_2+t_3+t_4)-t_2]\Delta G_2-\Delta G_3}{(t_2-t_1)(t_3-t_2)(t_4-t_2)}\\
\Delta\epsilon_3=\frac{[(t_1+t_2+t_3+t_4)-t_3]\Delta G_2-\Delta G_3}{(t_3-t_1)(t_3-t_2)(t_4-t_3)}\\
\Delta\epsilon_4=\frac{[(t_1+t_2+t_3+t_4)-t_4]\Delta G_2-\Delta G_3}{(t_4-t_1)(t_4-t_2)(t_4-t_3)}\\
\end{split}
\end{eqnarray}
Assuming $t_0=0$ and $t_4=1$ and vector $\Delta \vec{\epsilon}=(-1,2,-2,1)$ one finds:
\begin{eqnarray}
\begin{split}
&-t_2t_3=(t_2+t_3+1)\Delta G_2-\Delta G_3 \\
&-2t_2(t_3-t_2)(1-t_2)=(1+t_3)\Delta G_2-\Delta G_3\\
&-2t_3(t_3-t_2)(1-t_3)=(t_2+1)\Delta G_2-\Delta G_3\\
&-(1-t_2)(1-t_3)=(t_3+t_2)\Delta G_2-\Delta G_3
\end{split}
\end{eqnarray}
leading to:
\begin{eqnarray}
&t_2&=t_3-\frac{1}{2}\\
&t_3&
\end{eqnarray}
and
\begin{eqnarray}
\Delta G_2&=&-2\left(t_3-\frac{3}{4}\right)\\
\Delta G_3&=&-3\left(t_3-\frac{1-\sqrt{5}}{4}\right)\left(t_3-\frac{1+\sqrt{5}}{4}\right)
\end{eqnarray}

\subsubsection{case 1 :$\Delta G_2=0$}
\begin{equation}
 \lbrace{t_2=\frac{1}{4}t_4;t_3=\frac{3}{4}t_4; \Delta G_3=\frac{3}{16}t_{4}^{3}\rbrace}
\end{equation}
The solution leads to the identical double-loop interferometer insensitive to homogeneous acceleraton.
\subsubsection{case 2 : $\Delta G_3=0$}
\begin{equation}
 \lbrace{t_2=\frac{\sqrt{5}-1}{4};t_3=\frac{\sqrt{5}+1}{4};\Delta G_2=\left(1-\frac{\sqrt{5}}{2}\right)t_{4}^{2}\rbrace}
\end{equation}
The solution leads to a non-identical double-loop interferometer.

\section*{Acknowledgments}
The authors thank Marc Himbert (LCM-Cnam) and Michel Lefebvre (ONERA) for this fruitful collaboration between the two institutes.

\end{multicols}

\end{document}